# Measurement and Modeling on Terahertz Channels in Rain

Peian Li[1], Wenbo Liu[1], Jiacheng Liu[1], Da Li[1], Guohao Liu[1], Yuanshuai Lei[1], Jiabiao Zhao[1], Xiaopeng Wang[3], Jianjun Ma[1,2,*], John F. Federici[4]

[1]*Beijing Institute of Technology, Beijing, 100081, China*

[2]*Beijing Key Laboratory of Millimeter and Terahertz Wave Technology, Beijing Institute of Technology, Beijing, 100081, China*

[3]*Radar Operation Control Department, Beijing, 100044, China*

[4]*New Jersey Institute of Technology, New Jersey, 07102, USA*

*\*jianjun_ma@bit.edu.cn*

**Abstract**

The Terahertz (THz) frequency band offers a wide range of bandwidths, from tens to hundreds of gigahertz (GHz) and also supports data speeds of several terabits per second (Tbps). Because of this, maintaining THz channel reliability and efficiency in adverse weather conditions is crucial. Rain, in particular, disrupts THz channel propagation significantly and there is still lack of comprehensive investigations due to the involved experimental difficulties. This work explores how rain affects THz channel performance by conducting experiments in a rain emulation chamber and under actual rainy conditions outdoors. We focus on variables like rain intensity, raindrop size distribution (RDSD), and the channel's gradient height. We observe that the gradient height (for air-to-ground channel) can induce changes of the RDSD along the channel's path, impacting the precision of modeling efforts. To address this, we propose a theoretical model, integrating Mie scattering theory with considerations of channel's gradient height. Both our experimental and theoretical findings confirm this model's effectiveness in predicting THz channel behavior in rainy conditions. This work underscores the necessary in incorporating the variation of RDSD when THz channel travels in scenarios involving ground-to-air or air-to-ground communications.

## 1 Introduction

Terahertz (THz) communication offers a groundbreaking potential for achieving high data rates up to Tbps. However, the propagation losses suffered by THz waves, when transmitting through atmosphere, restrict their channel performances compared to conventional radio frequencies [1]. More specifically, the high sensitivity of THz waves to water makes wireless channels susceptible to weather conditions [2], such as rain, snow, and fog. This necessitates a deep understanding of atmospheric attenuation to build resilient THz communication setups [3].

Historically, initial investigations on the impact of rain on channel performance were primarily conducted using commercial microwave links [4, 5]. However, with the surging demand for high data rates and large bandwidths, research focus has transitioned towards millimeter-wave and subsequently THz communications [6-8]. Given that airborne particulates with sizes comparable to the THz wavelength can significantly affect the scattering, polarization, and extinction of propagating THz radiation [9], the attenuation due to rainfall is particularly significant, deeply influencing THz channel behavior.

Rain-induced channel performance degradation are commonly account for several factors, such as rain rate (i.e. rain intensity), raindrop size distribution (RDSD), shape of droplet and operating frequency. The RDSD is frequently characterized through mathematical models, including exponential [10], log-normal [11], and gamma distributions [12], *etc*. By comparing measured data with these models, researchers have been able to quantify rain-induced attenuation using empirical models like Mie scattering theory [13]. Illustratively, Hirata *et al*. compared their measured result at 120 GHz in rain with predictions by various RDSD models, and observed a notable alignment with the conditional Moupfouma distribution [14]. These results are often contrasted with models developed by the International Telecommunication Union (ITU). The ITU-R P.530, even though is used widely in rain weather conditions, presents regional limitations and miscalculated results in tropical climes with heavy rainfall patterns [15, 16]. This spatial variation in rainfall attenuation primarily emerges from disparities in rain rates and RDSD. Further, Wang *et al*. explored RDSD attributes under distinct rain rates and precipitation categories, emphasizing their influence on Gamma distribution parameters [17]. They found that, even within identical regions, raindrop dimensions and densities display obvious contrasts between dry and wet seasons [18].

Given the dynamic nature of rain, the efficiency of theoretical models are usually need to be optimized by employing outdoor measurements. However, factors, such as wind, can disturb and introduce inaccuracies in measurement. To mitigate such errors, Norouzian *et al.* discarded data recorded during wind speeds exceeding 10 m/s to prevent spurious numbers of drops during high wind events [19], while Weng *et al.* devised a theoretical model to assess wind-driven mechanical vibration and its impact on channel performance [20]. Federici *et al.* striving for a comprehensive understanding, developed weather chambers to emulate varied rain conditions, incorporating image analytic to discern RDSD and subsequently compute THz channel degradation by Mie scattering theory [21-23].

TABLE 1 SUMMARY OF PUBLICATIONS ON CHANNEL MEASUREMENT AND MODELING IN RAIN

| Frequency (GHz) | Distance | Scene | Remarks | Ref. |
| --- | --- | --- | --- | --- |
| 8-38 | 0.5-27 km | Outdoor | Commercial microwave links | [3] |
| 120 | 400 m | Outdoor | Moupfouma distribution, ITU model | [14] |
| 100-1000 | 4 m | Indoor | Mie scattering, Gaussian distribution | [21] |
| 625 | 3 m | Indoor | Weather emulating chamber | [22] |
| 14-39 | 1.5-7 km | Outdoor | Machine learning employed | [4] |
| 77-300 | 160 m | Outdoor | Corrected raindrop size distribution mode | [19] |
| 75-400 | 1-12 km | Outdoor | Violent rain, strong winds | [20] |
| 140-675 | 4 m | Indoor | QAM modulated data stream, Mie scattering | [23] |
| 25-77 | 36-200 m | Outdoor | Long term attenuation, ITU model | [29] |
| 130-150 | 70 m | Outdoor | Rain attenuation, multipath profile | [30] |
| 13-35 | 4-5 km | Outdoor | Backscattering model, vertical profile of rain | [31] |

Beyond terrestrial communication, rain-induced attenuation becomes even more pivotal in air-to-ground communication scenarios [24]. Along such a slant channel path with gradient height, the falling droplets undergo dynamic alterations in their density, dimensions, and falling velocity due to processes like coalescence and collisions [25]. Therefore, the vertical structure of rainfall characteristics is vital for optimizing wireless communication systems [34]. Ermis *et al.* formulated a backscattering model integrating particle swarm optimization with multilayered vector radiative transfer to deduce the vertical rain profile, though its efficacy failed at rain rates surpassing 12 mm/hr [31]. Most research efforts were more concentrated onto the vertical profile of rain, investigating factors such as RDSD, falling velocity, rain rate, liquid water content, and radar reflectivity profile, typically employing a Ka-band micro rain radar (MRR) [35, 36]. However, we still lack investigations on slant THz channels affected by gradient height-induced variation of RDSD, as indicated in Table 1. This obstructs the channel performance analysis for future air-to-ground or ground-to air communication paradigms.

This work tries to enrich the understanding of slant THz channel performance in rain by conducting controllable experimental measurements in laboratory. The influence caused by rain rate, RDSD, and channel's gradient height are measured and discussed. The subsequent sections are structured as follows: Section 2 details our experimental setup and generated rain droplets. Section 3 demonstrates the effect of RDSD on channel power attenuation. Section 4 discusses the performance of a slant channel in rain, and presents a theoretical model incorporating gradient height. Section 5 engages in air-to-ground communication simulations to validate the revised RDSD model's applicability. Section 6 presents outdoor measurements and modeling, and Section 7 encapsulates our discoveries and conclusions.

## 2 Experimental setup

The experimental setup developed for this work combines a THz transceiver system (similar to that in [23]) with a rain generation module (chamber), as illustrated in Fig. 1(a). The signal transmission component utilized a Ceyear 1465D signal generator with 0 dBm signal power, followed by a Ceyear 82406B frequency multiplier to achieve a terahertz frequency range of 110-170 GHz. The signal was then transmitted using an HD-1400SGAH25 horn antenna with a dielectric lens having a 10 cm focal length to enhance the gain of 25 dBi. The experimental results were demonstrated under a 140 GHz continuous wave signal. The receiving end mirrored this configuration, comprising an identical horn antenna, lens and a power sensor (Ceyear 71718).

The chamber is a polyethylene (PE) constructed module with dimensions of 17cm×11cm×5cm (length×width×height). It integrates a water inlet (with an inner diameter, $d$ = 3 mm) on one side and 98 water outlets (each having an inner diameter, $d$ = 2 mm) on the opposite end. For precise control over raindrop size, several kinds of medical needles (23G: $d_N$=0.6 mm, 21G: $d_N$=0.8 mm, 18G: $d_N$=1.1 mm) are affixed to these outlets. The inner diameter of the needles can determine the size of the generated raindrops directly. Within the chamber, the outlets cover an approximate area of 13cm×6cm (length × width), systematically arranged in a 14 × 7 grid pattern with a consistent 1 cm inter-outlet spacing. The chamber also includes a pressure pump and a flow meter, responsible for adjusting and maintaining the desired rainfall intensity. The entire chamber is fixed onto an optical platform using an iron frame. It's relative position and distance to the channel path (emulate the variation of channel height) can be changed by tilting a guide rail with the transmitter and receiver mounted on its both sides.

The rain model can provide stable rain rates ranging from 0 to 14296 mm/hr. To comprehensively study the rain-induced channel degradation, four distinct rain intensities were set with the rain rate ($Rr$) to be 733 mm/hr, 2786 mm/hr, 8065 mm/hr and 14296 mm/hr. It is noteworthy that such high intensities are unavailable in nature, but we think it is necessary for this work due to the restricted channel distance (< 1m) in our setup. Here, we choose both two highest rain intensities for further investigations. Considering the fact that RDSD usually determines the accuracy of channel modeling [37], we try to introduce the variation of raindrop size by employing three kind of needles with different diameters of 0.6 mm, 0.8 mm, and 1.1 mm. Fig, 1(b) shows that different raindrops can be achieved by our chamber under an identical rain intensity. The 0.6 mm inner diameter yields raindrops around 1-2 mm, the 0.8 mm inner diameter yields raindrops around 2-3 mm, and the 1.1mm inner diameter yields raindrops around 3-4 mm. This is necessary for us to analyze the influence of RDSD on our THz channel, while eliminating the disturbing of rain rate. It should be noted that there is a deviation of ~ ±0.5 mm for the droplet size, due to minor variations in the chamber's water pressure and collision coalescence, but we think this can be neglected in this work. The ambient temperature and humidity in our lab, which usually considered to be important factors affecting THz channel performance, can be controlled and kept fixed over the whole measurement (~25 °C and RH 42% ) by a air condition system.

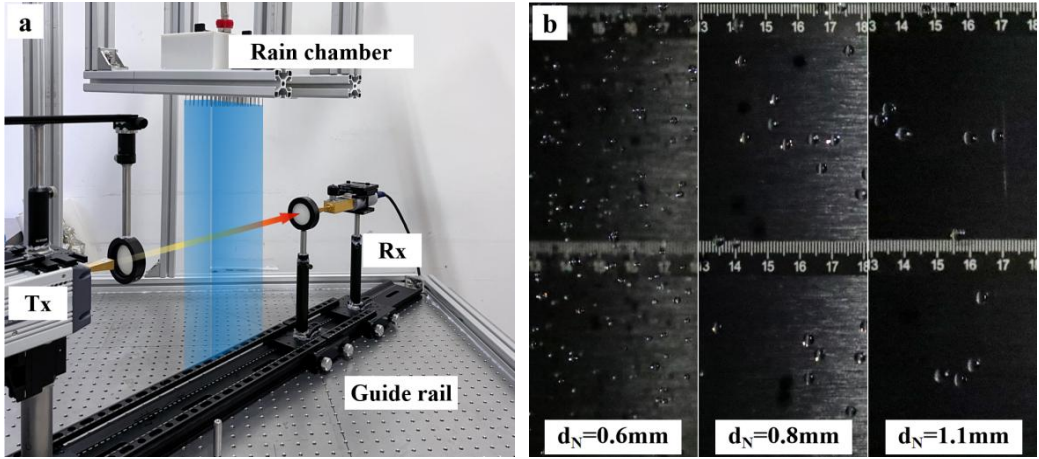

**Fig. 1** (a) Schematic diagram of the rainfall attenuation measurement setup. (b) Real shot of falling raindrops at high speed shutter.

## 3 Influence by raindrop size

Raindrops in the atmosphere typically exhibit sizes ranging from 0.1 mm to 8 mm. It is crucial to note that larger raindrops, due to the overpowering effects of aerodynamic forces compared to cohesive surface tension, do not sustain for extended durations in the atmosphere. The raindrop size distribution (RDSD) provides a metric for the number of drops per unit volume for each interval of drop diameter, while the rain rate offers an aggregated measure of rainfall depth over time. Although most rainfall phenomena correlate with the rain rate to some extent, it is conventional to represent RDSD as a function of this rate when designing theoretical models. However, it's imperative to understand that the density of raindrops per cubic meter can vary significantly. At the leading edge of a convective rainfall, the density may be sparse with only a few drops, while in denser regions, it can exceed 10,000 drops per cubic meter [25]. Thus, for accurate channel modeling, an in-depth analysis of RDSD under diverse conditions is critical.

Fig. 2 shows the measured channel power attenuation caused by the generated rain under an identical rain rate of 8065 mm/hr, with its raindrop size varies from 0.6 mm to 1.1 mm. A higher power attenuation is measured caused by the RDSD (0.6mm) with more rain droplets (smaller size) over the whole frequency range. This observation aligns with findings from reference [25], which emphasizes that THz channel degradation can differ substantially even if rain rates are identical - the distinction stems from the variation in raindrop sizes and numbers. This means that the impact of rain on THz channel is not only dependent on the rain intensity (rain rate) but also on the physical characteristics of the raindrops themselves, specifically their size and the total number of droplets.

It should also be note that the curves in Fig. 2(a) is not smooth, which is not due to our transceiver, but should be account to the power fluctuation caused by falling rain droplets. Fig. 2(b) shows the variation of power attenuation experienced by a 140 GHz channel over a time duration of 28 seconds. The measured data was recorded with sampling rate of 7 Hz. The inset shows the RMS (root mean square) of rainfall attenuation over this period, it is obvious that the RMS becomes smaller for larger raindrops with a smaller count number. We think this is due to that smaller raindrops with a larger count number increase the overall number passing through the channel path with reduced falling-velocity. This leads to a higher probability of interaction between the THz wave and the raindrops, resulting in more significant attenuation and scattering effects, and thereby higher power fluctuations.

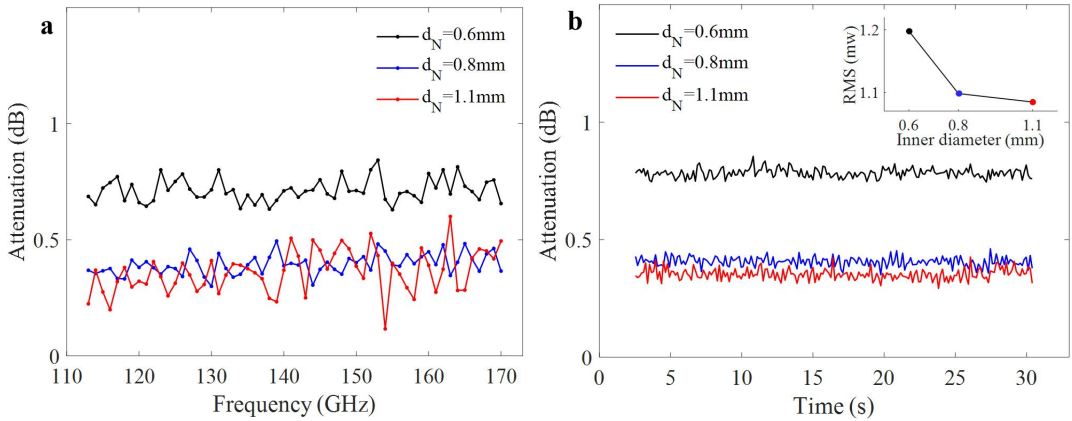

**Fig. 2** (a) Variation of rainfall attenuation with respect to operating frequency under different raindrop size distributions. (b) Rainfall attenuation on wireless channels operating at 140 GHz under a rain rate of 8065 mm/hr during 28s. Insert: Root mean square (RMS) of rainfall attenuation. For a 13 cm transmission in a region with a rainfall rate of 8065 mm/hr, a 0.3 dB attenuation corresponds to 2.3 dB/m, a 0.5 dB attenuation corresponds to 3.8 dB/m, a 0.6 dB attenuation corresponds to 4.6 dB/m, a 0.9 dB attenuation corresponds to 6.9 dB/m, and 1 dB attenuation corresponds to 7.7 dB/m.

## 4 Influence of rainfall height

It has been observed that the distribution of raindrops is altitude (height)-dependent [32, 34, 35], based on measurements of slant channels (with an elevation angle). However, there is no theoretical models contributing to this because there are many factor attributing to this phenomenon, such as variations of atmospheric pressure, temperature, and wind speed, which vary with respect to channel's height. It's worth noting that as raindrops falling, they encounter varying gravitational forces and aerodynamic drag based on their sizes. This dynamic interplay can lead to collisions leading to coalescence or fragmentation due to mechanical disruptions, altering the raindrop's microphysical attributes [25]. Consequently, these changes directly influence the RDSD at distinct heights. The processes of coalescence and fragmentation dynamically alter the RDSD as rain falls. This means that the RDSD at one height can be significantly different from the RDSD at another height due to these in-air transformations. For terrestrial communication channels, which operate at roughly the same height, the RDSD can be assumed to be uniform across the channel path. However, this assumption does not hold for air-to-ground or ground-to-air channels, where the channel path spans significantly different heights and, consequently, different RDSDs. Traditional models for predicting the impact of rain on THz channels often assume a fixed RDSD function over the entire channel path. This assumption simplifies the modeling process but fails to account for the dynamic nature of RDSD in the atmosphere, especially in air-to-ground communication scenarios. Such models might not accurately predict the attenuation and scattering effects caused by rain, as they do not consider the variation in RDSD with height. In this section, we try to emulate the variation of raindrop size distribution with respect to the distance between the rain chamber and the channel path. It is convenient that we can control the environment in our laboratory and extract the influence of wind, temperature change and up/down airflow, implying that the raindrop distribution may reach a form of stability. This makes the efficiency of our investigation on the influence of altitude on RDSD.

The experimental setup, as depicted in Fig. 3(a), was designed to measure the power attenuation under three different heights from the channel path to the experimental table. Here, we define the distances of channel height from the rain chamber as $H_{att}$, $M_{att}$, and $L_{att}$ with values of 51 cm, 105 cm and 153 cm, respectively. A smaller distance corresponds to a larger channel height, because the distance between the chamber and the experimental table is fixed. As the height changes, the count of raindrops intersecting it varies correspondingly as shown in Fig. 3(b), which equivalents to altitude variation of the THz channels. A high-speed camera with a shutter speed of 1/8000 seconds was deployed to capture these dynamics by taking numerous snapshots, registering instantaneous variations in raindrop counts within the channel path. In this section, we set the 0.8mm raindrop size for consideration. It is obvious that, for the distance of 51 cm, there are about 25 raindrops (average number) inside the channel beam in every moment. And this number reduces to approximately 16 and 11, for the distances of 105 cm, and 153 cm, respectively. A curve fitting was conducted by a power function as:

$$N_{rd} = 335.4 \times D^{-0.646} - 1.525 \tag{1}$$

where, $D$ (cm) is the distance between the rain chamber and the channel path, $N_{rd}$ represents the number of raindrops. Reference [25] mentions that the mass flux of raindrops is proportional to their final velocity. As raindrops fall, they collide and merge with one another, leading to an increase in their size. Additionally, due to air resistance, the velocity of a raindrop will eventually stabilize and reach what is known as its terminal velocity [26-28]. Based on this understanding, we can infer that in the region where the terminal velocity is stable, the size of the raindrops would also stabilize and hence, the number of raindrops would remain relatively constant. Therefore, in Fig. 2(b), we have utilized an power function to fit the data, because the initial part of this function aligns with our measured data, and it approaches a stable value in the latter half.

While rain rate offers a quantitative measure of ground-level rain intensity, it falls short in encapsulating the intricate microphysical characteristics of individual raindrops. To bridge this gap, we incorporated empirical mathematical models grounded on observed size spectra to emulate natural raindrop distributions. Considering the prevalence of larger raindrops during heavy rainfall and their significant sensitivity to altitude change [37], the diameter of generated raindrops in this experiment ranges from approximately 2 mm to 3mm, as indicated by the label "$d_N=0.8mm$" in Fig. 1(b). Extreme small or large raindrops were not considered, as emulating such a wide range of raindrop sizes is challenging in our laboratory. The raindrops generated by our setup predominantly conformed to an exponential distribution, described as:

$$N(D) = N_0 \exp(-\Lambda D) \tag{2}$$

where, $N(D)$ is the number density of raindrops with a diameter $D$ (mm) per unit volume, $N_0$ represents the number of raindrops within the radius $r+dr$ (mm) per unit volume, and $\Lambda$ characterizes the distribution's shape. $H_{att}$, $M_{att}$ and $L_{att}$ correspond to $N_0$ (mm$^{-1}$m$^{-3}$) of 450, 350, 250, and $\Lambda$ of $3R^{-0.21}$ mm$^{-1}$. Maintaining the premise that raindrops at various heights, but from a uniform source, follow a consistent distribution pattern, we modified the $N_0$ parameter according to our measurements. Subsequently, using the Mie scattering theory, we computed the rainfall attenuation. Our model's predictions exhibited remarkable concordance with empirical data, as portrayed in Fig. 3(c).

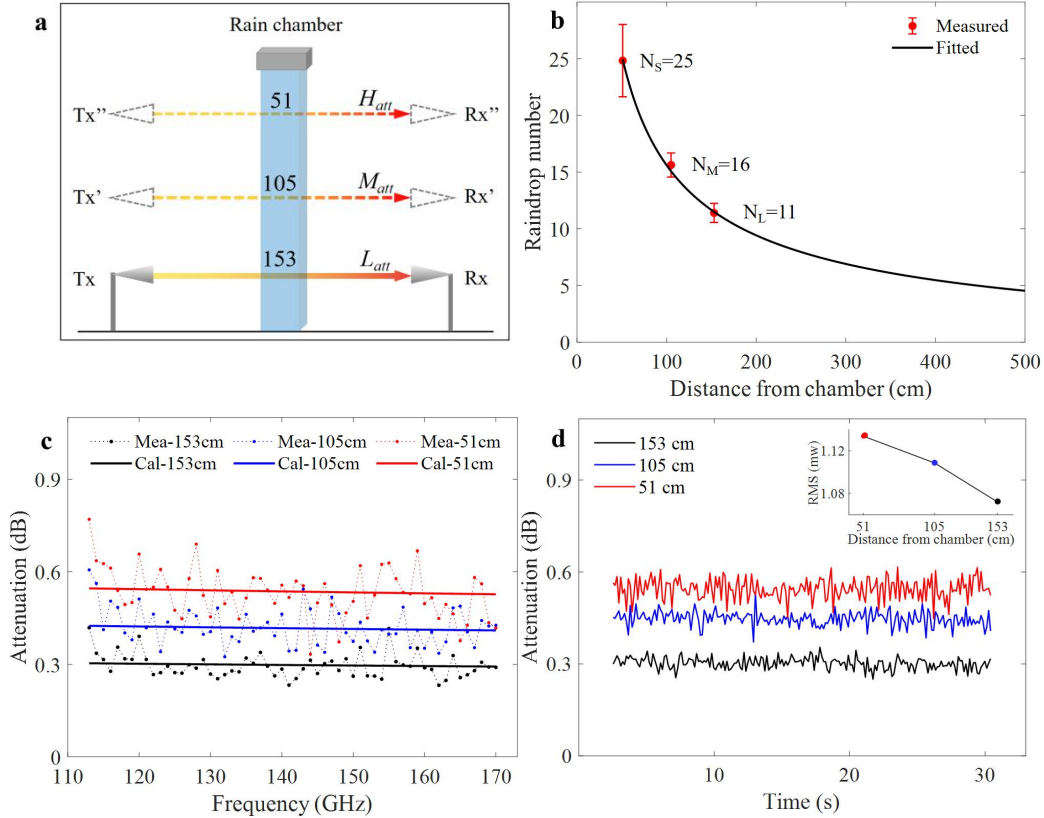

**Fig. 3** (a) Illustration detailing signal propagation across varying heights within the rain-affected zone (specific beam distances from the chamber are 51 cm, 105 cm, and 153 cm). (b) Instantaneous fluctuations in the count of raindrops traversing the channel beam at distinct heights. (c) Comparative data on empirically observed and theoretically computed rain attenuation, stratified by operating frequencies and heights. (d) Analysis of attenuation in wireless channels at 140GHz, subjected to a rain intensity of 8065 mm/hr over a span of 28 seconds. Inset: Root Mean Square (RMS) analysis of rainfall-induced attenuation across the three examined heights (51cm, 105cm, 153cm).

The signal perturbations at 140 GHz were monitored over a designated period, with their root mean square (RMS) subsequently computed. Fig. 3(d) shows that as the beam approaches the rain chamber, there's a noticeable upsurge in rainfall attenuation, culminating in heightened signal perturbations. In situations where raindrop sizes remain relatively comparable, the count of raindrops within the beam and the inherent randomness they introduce predominantly dictate signal stability.

**5 Influence of gradient height**

Communication services that operate from air to ground, such as aviation communication systems, meteorological observation, necessitate both high reliability and expansive bandwidth. Among the plethora of variables affecting these systems, the alteration of raindrop size distribution due to altitude (channel height) shifts stands out as particularly significant [33].

To empirically validate the implications of this phenomenon, we carried out controlled laboratory condition to emulate the air-to-ground channel (slant channel with gradient height) at 140GHz with the rain rate of 14296 mm/hr. As demonstrated in Fig. 4(a), we fixed the transmitter and receiver at both sides of the guiding rail, ensuring a constant channel distance. The measurement data is presented in Fig. 4(b) and we can find that an increase in the elevation angle (related to height gradient) leads to more serious attenuation. Our initial focus lay on assessing the rain-induced attenuation based on the alteration in path due to shifting elevation angles. But the path-induced attenuation (dotted line in Fig. 4(b)) shows an obvious lower value compared to the measurement, a more intricate interpretation of the acquired data is necessary due to this significant difference.

Incorporating our updated model of raindrop size distribution, which now includes in gradient height changes, we introduced three parameters $L_{att}$, $M_{att}$, and $H_{att}$ in Fig. 3(a), to represent the power attenuation for distances of 153cm, 105cm, and 51cm, respectively. As shown in Fig. 4(b), at an elevation angle of 0°, where the channel path is parallel to the ground, only $L_{att}$ is applicable without height-induced factor. Yet, with the elevation angle increases, both $M_{att}$ and $H_{att}$ values show a noticeable increasing, while $L_{att}$ exhibits a decline in proportion. To provide a clearer understanding of height-induced power attenuation in a slant rainfall channel, we introduced a mathematical equation as shown by the solid lines in Fig. 4(b):

$$PL = \frac{H_m}{H_a}(\alpha \cdot L_{att} + \beta \cdot M_{att} + \gamma \cdot H_{att}) \qquad (3)$$

where, $PL$ signifies the path loss of the signal. $H_m$ represents the vertical distance of the Rx experiences when the elevation angle changes from 0° to the desired value. In this work, the maximum elevation angle is around 45°, which is limited by our setup. The parameter $H_a$ accounts for the actual channel height. The ratio represents the increase/decrease factor. We divided the distance between the chamber and the channel path to three parts in this work, with each part presumed to have a constant signal attenuation corresponding to $L_{att}$, $M_{att}$ and $H_{att}$. The proportionality constants $\alpha$, $\beta$ and $r$ pertain to these divisions, symbolizing the fractions of the signal traversing the rain-affected region at diverse elevation angles. While our current model operates on these three divisions, subsequent inquiries could introduce more nuanced segmentation tailored to specific use-cases, enhancing the accuracy of this model. Fig. 4(c) pictorially delineates the computed and adapted values for these proportion factors. As elevation augments, the influence of $M_{att}$ and $H_{att}$ strengthens, contrasted by a decrease in $L_{att}$'s contribution.

Signal attenuation fluctuates with different RDSDs (height), where heightened attenuation correlates with the presence of larger or more raindrops [36].

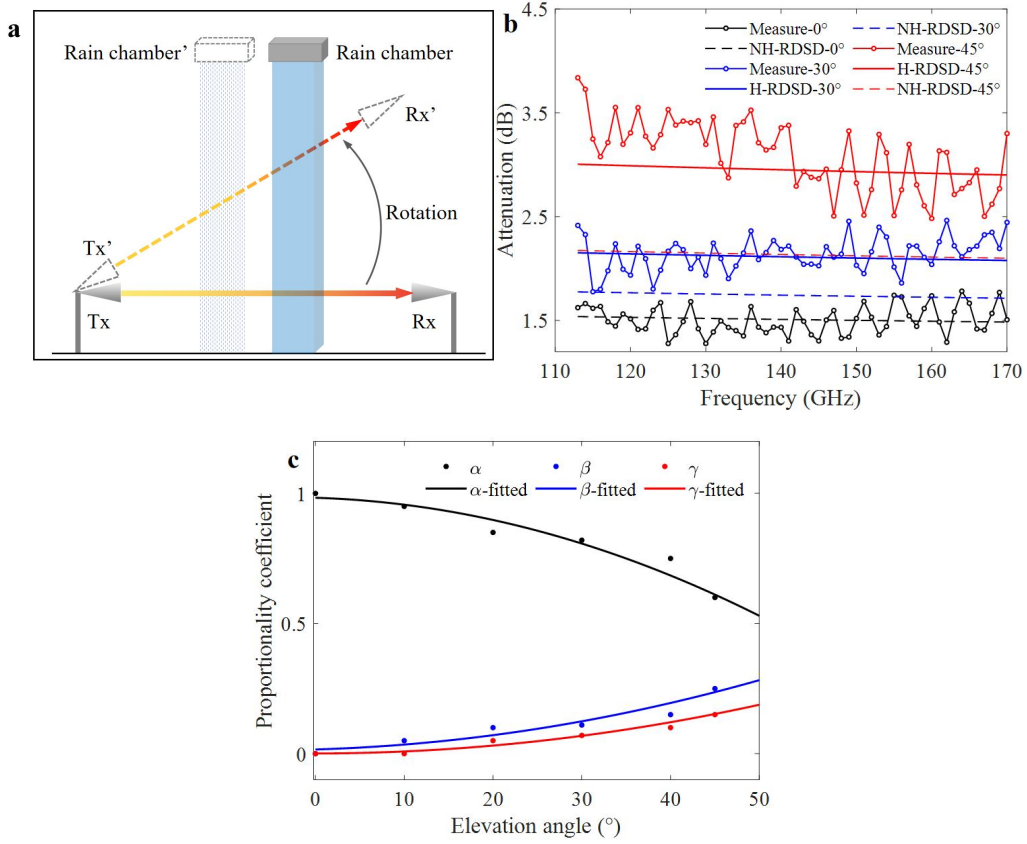

**Fig. 4** (a) Illustration of signal propagation across varied elevation angles, spanning 0° to 45°. (b) Measured and predicted results with/without the height-induced raindrop size alterations at 140GHz with the rain rate of 14296 mm/hr. The dotted line represents signal attenuation considering elevation-induced rainfall path changes, whereas the solid lines take into account additional signal attenuation incorporating height-induced RDSD. The color-coding in the graph aligns with that in Fig. 4(c). (c) Computed and fitted metrics for the three proportionality constants. Point values signify computed data, whereas continuous lines represent the adjusted values. The black data set ($\alpha$ / $\alpha$-fitted) is indicative of the proportionality factor when the rain chamber stands at a distance of 153cm from the beam, blue data ($\beta$ / $\beta$-fitted) pertains to a height of 105cm, and red data ($\gamma$ / $\gamma$-fitted) is associated with a 51cm elevation.

## 6 Outdoor experimental validation

A persistent challenge in understanding rain-induced attenuation in communication channels lies in the constraints of empirical models, which predominantly hinge on limited-range rainfall rate measurements. Such models might not fully account for the nuanced influence of factors such as spatial structures, height-related changes in RDSD, and varied climatic conditions. Recognizing this gap, this work introduces a correction model that accounts for height-induced RDSD variations above.

In addition, drastic changes in rain rates can also cause variations in RDSD, yet many traditional models are designed for specific ranges of rainfall intensity. To put our RDSD-based correction model to the test and establish its viability, we embarked on a systematic outdoor experimentation phase. The experimental setup was strategically located in the Liangxiang campus of the Beijing Institute of
9

Technology. This location provided us with a 41.4-meter-long communication channel nestled between two infrastructural edifices, as vividly illustrated in Fig. 5(a). Our methodology was comprehensive: measuring channel performance not only under rainy conditions but also when the skies were clear, offering a benchmark comparison.

The experiment was timed with the onset of Typhoon "DuSuri", a circumstance that enriched our dataset with a wide spectrum of rainfall rates. Systematic data acquisition was paramount: signal data was meticulously logged at one-minute intervals spanning a duration of 90 minutes. Simultaneously, rain rate recordings were taken every ten minutes. A scrutiny of Fig. 5(b) showcases the distribution of the captured rain rates, with pronounced concentrations at 6.8 mm/hr, 7.6 mm/hr, 15.3 mm/hr, and 30.6mm/hr.

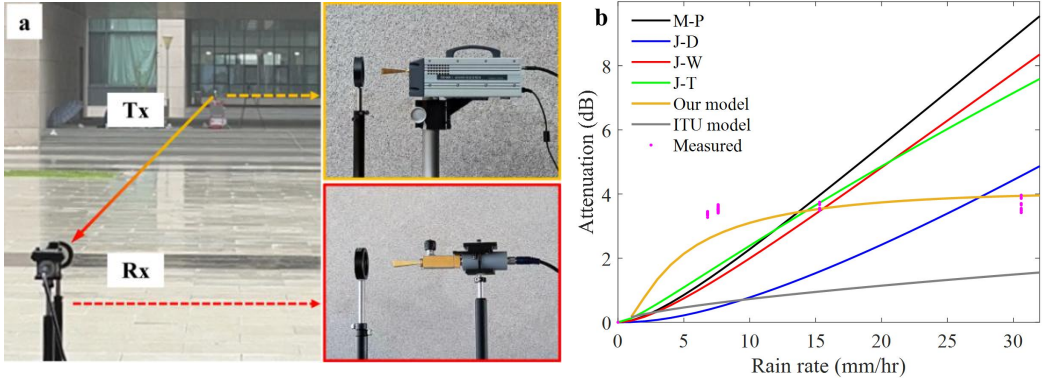

**Fig. 5** (a) Schematic representation of the outdoor setup designed to gauge rainfall attenuation experienced by a 140 GHz channel. (b) Calculations from prevailing models (M-P, J-D, J-W, J-T [14, 38]) and ITU model [39] with those from our innovative RDSD correction model, juxtaposed with real-time measurements spanning diverse rainfall intensities.

Building upon the widely-accepted exponential distribution model, we incorporated refinements by adjusting both its quantity and shape parameters in accordance with the real-time observed rain rates. Rain rate of 6.8 mm/hr, 7.6 mm/hr, 15.3 mm/hr, 30.6 mm/hr correspond to $N_0$ (mm$^{-1}$m$^{-3}$) of 1850, 1680, 1000, 620, and $\Lambda$ of $4R^{-0.21}$ mm$^{-1}$. Utilizing these data, we obtained the $N_0$ function that varies with rainfall rate R while maintaining $\Lambda$ constant, to derive the RDSD model for rain rate variation. To ascertain rainfall attenuation, the Mie scattering theory was integrated, and the subsequent results were elegantly encapsulated in the calculation curve represented by the yellow line in Fig. 5(b). Encouragingly, compared to the results of the ITU-R p.838-3 model [39] and other RDSD models, our findings demonstrated commendable alignment with the empirically measured values, thereby reinforcing the credibility and utility of our proposed RDSD correction model. The parameters of other RDSD models are shown in Table 2. Our model can work for a broader range of rain rates, but given the highly complex nature of rainfall, further optimization is necessary when conditions change.

TABLE 2 PARAMETERS OF RAINDROP SIZE DISTRIBUTION

| RDSD model | $N_0$ (mm$^{-3}$mm$^{-1}$) | $\Lambda$ (mm$^{-1}$) |
| --- | --- | --- |
| M-P [14] | 8000 | $4.1R^{-0.21}$ |
| J-D [14] | 30000 | $5.7R^{-0.21}$ |
| J-W [38] | 7000 | $4.1R^{-0.21}$ |
| J-T [38] | 1400 | $3.0R^{-0.21}$ |

# 7 Conclusion

This work investigates the impact of rain on THz channels, a critical endeavor given the THz frequency band's promise for ultra-high-speed data transmission. By examining variables such as rain intensity, RDSD, and the channel's gradient height through both indoor rain emulation and outdoor experiments, we unveils the influence of rain, especially its variable RDSD influenced by the channel's gradient height, on THz channel performance. Our findings underscore the insight that the elevation angle of THz channels, particularly in air-to-ground communications, introduces variability in RDSD along the channel's path, thereby impacting the accuracy of channel models. We propose a novel theoretical model that integrates Mie scattering theory with considerations of the channel's gradient height, which our experimental and theoretical results validate as effective in predicting THz channel behavior under rainy conditions. We think this work reinforces the necessity of incorporating RDSD variability in THz channel models, especially for scenarios involving gradient heights, to enhance the reliability and efficiency of THz communications in adverse weather conditions.

**Author Contribution**

**Peian Li**: Methodology, Experiment, Calculation, Validation, Investigation, Writing. **Wenbo Liu**: Design, Methodology, Experiment, Investigation. **Jiacheng Liu**: Experiment. **Da Li**: Experiment. **Guohao Liu**: Experiment. **Yuanshuai Lei**: Experiment. **Jiabiao Zhao**: Investigation. **Xiaopeng Wang**: Meteorological data provision. **Houjun Sun**: Investigation, Validation. **Jianjun Ma**: Conceptualization, Methodology, Experiment, Funding acquisition, Writing, Supervision. **John F. Federici**: Writing, Supervision.

**Funding**

This work was supported in part by the National Natural Science Foundation of China (62071046, 92167204, 62072030), the Science and Technology Innovation Program of Beijing Institute of Technology (2022CX01023), the Graduate Innovative Practice Project of Tangshan Research Institute, BIT (TSDZXX202201) and the Talent Support Program of Beijing Institute of Technology "Special Young Scholars" (3050011182153).

**Data Availability**

Data underlying the results presented in this paper are not publicly available at this time but may be obtained from the authors upon reasonable request.

**Declarations**

Ethics Approval: Not applicable.
**Consent for Publication:** All the authors agreed to publish.
**Conflict of Interest:** The authors declare no competing interests.

Reference

[1] R. Wang, Y. Mei, X. Meng, and J. Ma, "Secrecy performance of terahertz wireless links in rain and snow," *Nano Communication Networks,* vol. 28, p. 100350, 2021. https://doi.org/10.1016/j.nancom.2021.100350

[2] J. Lou et al. "Calibration-free, high-precision, and robust terahertz ultrafast metasurfaces for monitoring gastric cancers", Proceedings of the National Academy of Sciences of the United States of America, vol. 119, no. 43, pp. e2209218119, 2022. https://doi.org/10.1073/pnas.2209218119

[3] A. Zinevich, P. Alpert, and H. Messer, "Estimation of rainfall fields using commercial microwave communication networks of variable density," Advances in water resources, vol. 31, no. 11, pp. 1470-1480, 2008. https://doi.org/10.1016/j.advwatres.2008.03.003

[4] D. Jacoby, J. Ostrometzky, and H. Messer, "Short-term prediction of the attenuation in a commercial microwave link using LSTM-based RNN," in 2020 28th European Signal Processing Conference (EUSIPCO), 2021: IEEE, pp. 1628-1632. https://doi.org/10.23919/Eusipco47968.2020.9287835

[5] E. Alozie et al., "A review on rain signal attenuation modeling, analysis and validation techniques: Advances, challenges and future direction," Sustainability, vol. 14, no. 18, p. 11744, 2022. https://doi.org/10.3390/su141811744

[6] A. M. Al-Saman, M. Cheffena, M. Mohamed, M. H. Azmi, and Y. Ai, "Statistical analysis of rain at millimeter waves in tropical area," IEEE Access, vol. 8, pp. 51044-51061, 2020. https://doi.org/10.1109/ACCESS.2020.2979683

[7] L. Luini, G. Roveda, M. Zaffaroni, et al. "The Impact of Rain on Short E -Band Radio Links for 5G Mobile Systems: Experimental Results and Prediction Models," IEEE Transactions on Antennas and Propagation, vol. 68, no. 4, pp.3124-3134, 2019. https://doi.org/10.1109/TAP.2019.2957116

[8] J. Huang, Y. Cao, X. Raimundo, et al. "Rain statistics investigation and rain attenuation modeling for millimeter wave short-range fixed links," IEEE Access, vol. 7, pp. 156110-156120, 2019. https://doi.org/10.1109/ACCESS.2019.2949437

[9] A. Bandyopadhyay, et al. "Effects of scattering on THz spectra of granular solids." International Journal of Infrared and Millimeter Waves, vol. 28, pp. 969-978, 2007. https://doi.org/10.1007/s10762-007-9276-y

[10] J. S. Marshall and W. M. K. Palmer, "The distribution of raindrops with size," Journal of Atmospheric Sciences, vol. 5, no. 4, pp. 165-166, 1948.

[11] D. A. de Wolf, "On the Laws-Parsons distribution of raindrop sizes," Radio Science, vol. 36, no. 4, pp. 639-642, 2001. https://doi.org/10.1029/2000RS002369

[12] G. O. Ajayi and R. L. Olsen, "Modeling of a tropical raindrop size distribution for microwave and millimeter wave applications," Radio Science, vol. 20, no. 02, pp. 193-202, 1985. https://doi.org/10.1029/RS020i002p00193

[13] C. Mätzler, "Drop-size distributions and Mie computations for rain," 2002.

[14] A. Hirata et al., "Effect of rain attenuation for a 10-Gb/s 120-GHz-band millimeter-wave wireless link," IEEE transactions on microwave theory and techniques, vol. 57, no. 12, pp. 3099-3105, 2009. https://doi.org/10.1029/RS020i002p00193

[15] R. Series, "Propagation data and prediction methods required for the design of terrestrial line-of-sight systems," Recommendation ITU-R, p. 530-18, 2021.

[16] O. Darley, A. I. Yussuff, and A. A. Adenowo, "Investigation into rain attenuation prediction models at locations in lagos using remote sensing," FUOYE Journal of Engineering and Technology, vol. 6, no. 2, pp. 19-24, 2021.

[17] G. Wang, R. Li, J. Sun, X. Xu, R. Zhou, and L. Liu, "Comparative analysis of the characteristics of rainy season raindrop size distributions in two typical regions of the Tibetan Plateau," Advances in Atmospheric Sciences, vol. 39, no. 7, pp. 1062-1078, 2022. https://doi.org/10.1007/s00376-021-1135-6

[18] R. Lai et al., "Raindrop size distribution characteristic differences during the dry and wet seasons in South China," Atmospheric Research, vol. 266, p. 105947, 2022. https://doi.org/10.1016/j.atmosres.2021.105947

[19] F. Norouzian et al., "Rain attenuation at millimeter wave and low-THz frequencies," IEEE Transactions on Antennas and Propagation, vol. 68, no. 1, pp. 421-431, 2019. https://doi.org/10.1109/TAP.2019.2938735

[20] Z.-K. Weng et al., "Millimeter-wave and terahertz fixed wireless link budget evaluation for extreme weather conditions," IEEE Access, vol. 9, pp. 163476-163491, 2021. https://doi.org/10.1109/ACCESS.2021.3132097

[21] J. Ma, F. Vorrius, L. Lamb, L. Moeller, and J. F. Federici, "Comparison of experimental and theoretical determined terahertz attenuation in controlled rain," Journal of Infrared, Millimeter, and Terahertz Waves, vol. 36, pp. 1195-1202, 2015. https://doi.org/10.1007/s10762-015-0200-6

[22] J. Ma, F. Vorrius, L. Lamb, L. Moeller, and J. F. Federici, "Experimental comparison of terahertz and infrared signaling in laboratory-controlled rain," Journal of Infrared, Millimeter, and Terahertz Waves, vol. 36, pp. 856-865, 2015. https://doi.org/10.1007/s10762-015-0183-3

[23] P. Li et al., "Performance degradation of terahertz channels in emulated rain," Nano Communication Networks, vol. 35, p. 100431, 2023. https://doi.org/10.1016/j.nancom.2022.100431

[24] V. Doborshchuk, V. Begishev, and K. Samouylov, "Propagation Model for Ground-to-Aircraft Communications in the Terahertz Band with Cloud Impairments," *Energies,* vol. 15, no. 21, p. 8022, 2022. https://doi.org/10.3390/en15218022

[25] G. Zhang, "Dual polarization radar meteorology," China Meteorological Press, 2018.

[26] D. Atlas, C. W. Ulbrich, "Path-and area-integrated rainfall measurement by microwave attenuation in the 1-3 cm band," Journal of Applied Meteorology, vol: 16, pp. 1322-1331, 1977. https://doi.org/10.1175/1520-0450(1977)016<1322:PAAIRM>2.0.CO;2

[27] D. Atlas, R. C. Srivastava, and R. S. Sekhon, "Doppler radar characteristics of precipitation at vertical incidence," Reviews of Geophysics, vol. 11, pp. 1-35, 1973. https://doi.org/10.1029/RG011i001p00001

[28] E. A.Brandes, G Zhang, and J. Vivekanandan, "Experiments in rainfall estimation with a polarimetric radar in a subtropical environment," Journal of Applied Meteorology, vol. 41, pp. 674-685, 2022. https://doi.org/10.1175/1520-0450(2002)041<0674:EIREWA>2.0.CO;2

[29] O. Zahid and S. Salous, "Long-term rain attenuation measurement for short-range mmWave fixed link using DSD and ITU-R prediction models," Radio Science, vol. 57, no. 4, pp. 1-21, 2022. https://doi.org/10.1029/2021RS007307

[30] P. Sen et al., "Terahertz communications can work in rain and snow: Impact of adverse weather conditions on channels at 140 GHz," in Proceedings of the 6th ACM Workshop on Millimeter-Wave and Terahertz Networks and Sensing Systems, 2022, pp. 13-18.

[31] S. Ermis and E. Yigit, "Estimation of rain parameters for microwave backscattering model using PSO," Atmospheric Research, vol. 288, p. 106720, 2023. https://doi.org/10.1016/j.atmosres.2023.106720

[32] H. Zheng, Y. Zhang, L. Zhang, H. Lei, and Z. Wu, "Precipitation microphysical processes in the inner rainband of tropical cyclone Kajiki (2019) over the South China Sea revealed by polarimetric radar," Advances in Atmospheric Sciences, vol. 38, pp. 65-80, 2021. https://doi.org/10.1007/s00376-020-0179-3

[33] A. M. Shalaby and N. S. Othman, "The effect of rainfall on the uav placement for 5g spectrum in malaysia," Electronics, vol. 11, no. 5, pp. 681, 2022. https://doi.org/10.3390/electronics11050681

[34] S. Das, and A. Maitra. "Vertical profile of rain: Ka band radar observations at tropical locations," Journal of Hydrology, vol. 534, pp. 31-41, 2016. https://doi.org/10.1016/j.jhydrol.2015.12.053

[35] R. Ramadhan, Marzuki, M. Vonnisa, et al. "Diurnal Variation in the Vertical Profile of the Raindrop Size Distribution for Stratiform Rain as Inferred from Micro Rain Radar Observations in Sumatra," Advances in Atmospheric Sciences, vol. 37, pp. 832-846, 2020. https://doi.org/10.1007/s00376-020-9176-9

[36] C. Song, Y. Zhou, and Z. Wu. "Vertical profiles of raindrop size distribution observed by micro rain radar," Appl Meteor Sci, vol. 30, no. 4, pp. 479-490, 2019.

[37] F. Norouzian, E. Marchetti, Gashinova, et al. "Rain attenuation at millimeter wave and low-THz frequencies," IEEE Transactions on Antennas and Propagation, vol, 68, no. 1, pp. 421-431. https://doi.org/10.1109/TAP.2019.2938735.

[38] J. Joss, J. C. Thams, A, Waldvogel. "The variation of raindrop size distributions at Locarno," proc.int.conf.cloud physics, 1968.

[39] R. Series, "Specific attenuation model for rain for use in prediction methods," Recommendation ITU-R, p. 838-3, 2005.